\documentclass[12pt,preprint]{aastex}

\begin{document}

\title{$r$-Process Nucleosynthesis in Hot Accretion Disk Flows from Black Hole - Neutron 
Star Mergers}

\author{R. Surman\altaffilmark{1} and G. C. McLaughlin}
\affil{Department of Physics, North Carolina State University, Raleigh, NC 27695-8202}

\author{M. Ruffert}
\affil{School of Mathematics, University of Edinburgh, Edinburgh EH9 3JZ}

\author{H.-Th. Janka}
\affil{Max-Planck-Institut f{\"u}r Astrophysik, Postfach 1317, 85741 Garching, Germany}
\and 
\author{W. R. Hix}
\affil{Physics Division, Oak Ridge National Laboratory, Oak Ridge, TN 37831-6374}

\altaffiltext{1}{Permanent Address:  Department of Physics and Astronomy, Union College, Schenectady, NY 12308  }

\begin{abstract}

We consider hot accretion disk outflows from black hole - neutron star mergers in the context of the
nucleosynthesis they produce.  We begin with a three dimensional numerical model of a black hole -
neutron star merger and calculate the neutrino and antineutrino fluxes emitted from the resulting
accretion disk.  We then follow the element synthesis in material outflowing the disk along
parameterized trajectories.  We find that at least a weak $r$-process is produced, and in some cases a
main $r$-process as well.  The neutron-rich conditions required for this production of $r$-process
nuclei stem directly from the interactions of the neutrinos emitted by the disk with the free neutrons
and protons in the outflow.

\end{abstract}


\section{Introduction}

Compact object mergers, such as neutron star - neutron star (NS-NS) mergers or black hole -
neutron star (BH-NS) mergers, have long been speculated to be a site of the $r$-process of
nucleosynthesis \citep{lat74,lat76}.  There are a number of different processes which can
create ejecta during a merger.  For both NS-NS and BH-NS mergers, cold, low entropy neutron
star matter originating from the inner crust of the merging neutron star(s) can be ejected
from tidal tails which form during the merger.  Traditional calculations of nucleosynthesis
from mergers have focused on this mechanism by calculating either cold decompression of
neutron star matter, e.g. \citep{mey89,gor05} or mild heating of the matter to $T \sim 1$
MeV which leads to an $(n,\gamma)$-$(\gamma,n)$ equilibrium, e.g. \citep{fre99}.  Recently,
\citet{oec07} have shown that in addition to the ejection from tidal tails, in the case of
NS-NS mergers there is a second, hotter ($T\sim 10$ MeV) component that comes from the
contact surface of the two merging neutron stars and is ejected perpendicular to the plane
of the merger.  As either a NS-NS or BH-NS merger progresses, an accretion torus surrounding
a black hole is formed.  Additional material can be ejected from the inner, hot region of
the torus and/or the colder outer region, either out of the plane of the disk or escaping
from the edges of the torus.  Little calculation has been performed to date on the latter
scenarios; we will focus here on ejecta from the inner hot region of the torus.

Understanding of the $r$-process of nucleosynthesis has undergone a dramatic transition in
the last decade.  Prior to this transition, it was a commonly held view that there was one
primary process that contributed to the entire range of $r$-process nuclei; recent data from
meteorites and metal-poor halo stars has changed that view (see \citet{qian07} and
references therein). This data independently suggests that there is a main $r$-process which
contributes to the heavier $r$-process elements above the second peak and a weak $r$-process
that contributes to those below.  Using timescale arguments based on the data further
suggests that compact object mergers are not primarily responsible for the main $r$-process
component seen in the metal poor halo stars \citep{arg}, though a contribution may be
possible if a second production mechanism for compact binaries exists that results in the
creation of tighter orbits with faster decay timescales, as has been suggested by
\citet{bel06}.  In any event compact binary mergers could possibly be a significant
contributor to the weak $r$-process component as observed in the solar system today.

Interest in BH-NS mergers has recently intensified with the suggested correlation with short duration
gamma ray bursts, e.g., \citet{ruf97,janka99,nak07} and references therein.  State of the art numerical
models now include high-resolution shock-capturing hydrodynamics (on nested grids or with sufficiently
high particle numbers) including some amount of relativistic effects and the ability to trace the nuclear
composition, a matching microphysical equation-of-state, and including neutrino emission and its main
effects on the emitting matter.  This creates an excellent launching point for studies of
nucleosynthesis.

In this letter we present the results of nucleosynthesis calculations for material in winds
which leave the surface of the hot accretion disk formed by the merger of a black hole and a
neutron star.  We take special care with the neutrino fluxes, since neutrino interactions
primarily set the electron fraction ($Y_e = {p \over {n + p}}$), and therefore determine the 
outcome of the nucleosynthesis.

\section{Black Hole - Neutron Star Merger Model}

We simulate the merger of a non-rotating cool $1.6 M_\odot$ neutron star and a $2.5 M_\odot$ 
black hole with spin parameter 0.6. The hydrodynamics is evolved with the Piecewise 
Parabolic Method on grids with $128^3$ zones nested 4 deep \citep{ruf96}. Gravitational 
wave emission and neutrino emission is included as described in Ruffert \& Janka
(2001, and references therein). The black hole is treated as gravitating vacuum sphere with 
a modified potential \citep{art96}. The gas is described by equation of state of 
\citet{shen98}. The simulation was evolved until the remains of the shredded neutron
star formed a disk around the black hole, i.e.~approximately 20 ms after initial contact.

\begin{figure}
\plotone{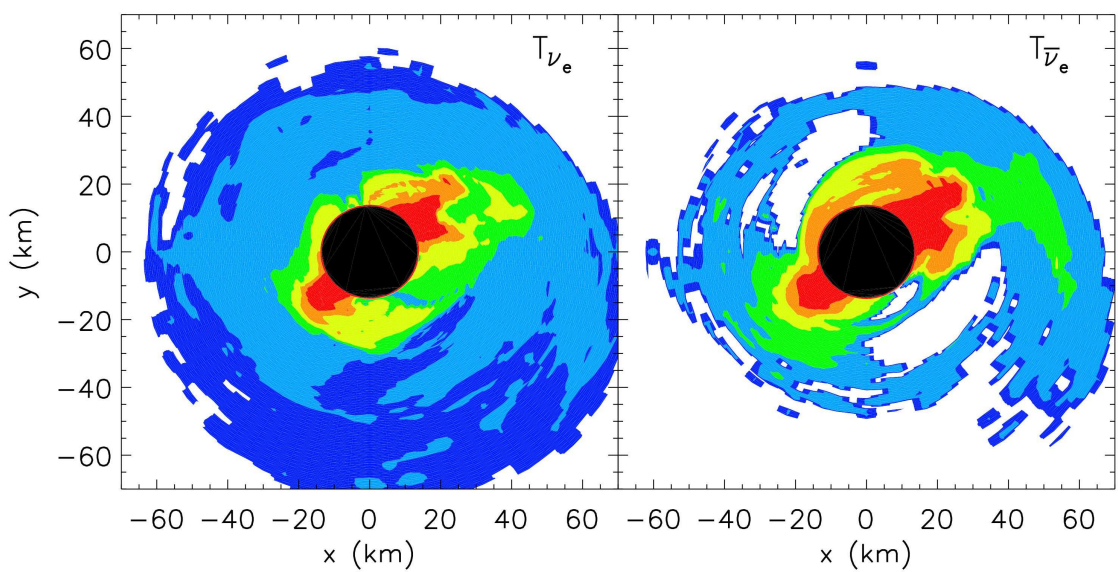}
\caption{Plots of the electron neutrino (left panel) and electron antineutrino (right panel)
temperature from the surface of the disk.  Only regions where the neutrinos are trapped are
shown. One can see from the figure that disks are asymmetric.  Contours indicate regions,
from darkest to lightest (or blue to red in color version), of temperatures of 1 MeV, 3 MeV,
5 MeV, 7 MeV, 9 MeV, 11 MeV and 13 MeV.  The dark center indicates the inner boundary of 
the numerical merger model.
\label{fig:nusph}}
\end{figure}

\begin{figure}
\plotone{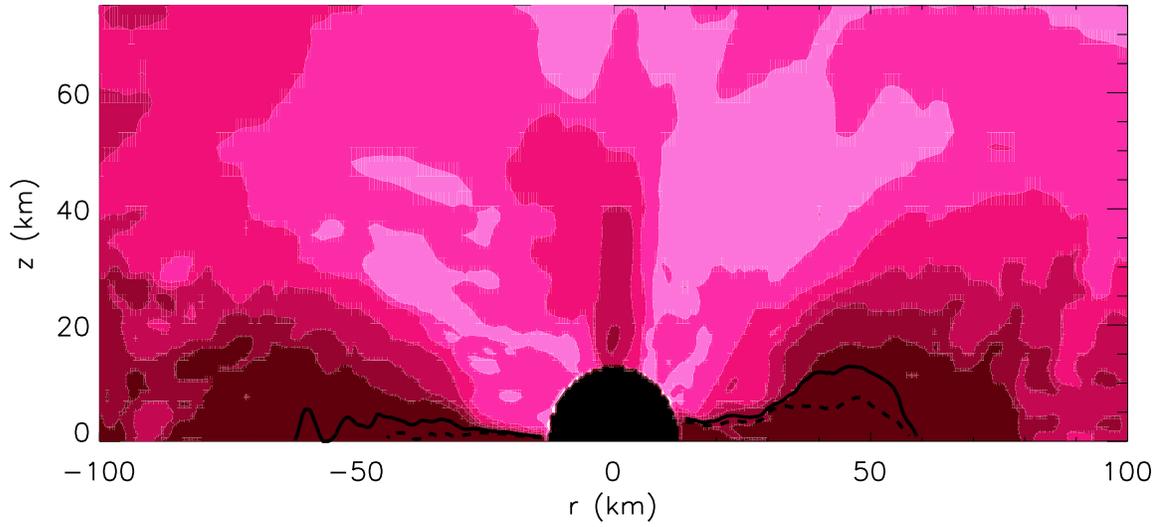}
\caption{Shows density along a vertical slice of the disk. The shaded regions, from lightest
to darkest, show densities of $10^{8.5}$, $10^9$, $10^{9.5}$, $10^{10}$, $10^{10.5}$, and
$10^{11}$ g/cm$^3$.  The solid line shows the electron neutrino surface while the dashed
line shows the electron antineutrino surface.  The dark center indicates the inner boundary
of the numerical merger model.
\label{fig:vertical}}
\end{figure}

From this model we have calculated neutrino surfaces, i.e.\ the places where the neutrinos
decouple from the disk, using the method described in \citet{sur04}. The results can be seen
in Fig.\ \ref{fig:nusph} where we show the local disk temperatures at the point of
decoupling, which we take to be the temperatures of thermal neutrino fluxes emitted from
the disk.  One can see that the spatial region of neutrino trapping exceeds that of
antineutrino trapping but that the temperatures of the emitted $\bar{\nu}_e$ are greater
than that of the emitted $\nu_e$.  As we will show, this has important consequences for the
nucleosynthesis.  We also show a vertical slice of the disk in Fig.\ \ref{fig:vertical}.

\section{Nucleosynthesis Calculation}

\begin{figure}
\plotone{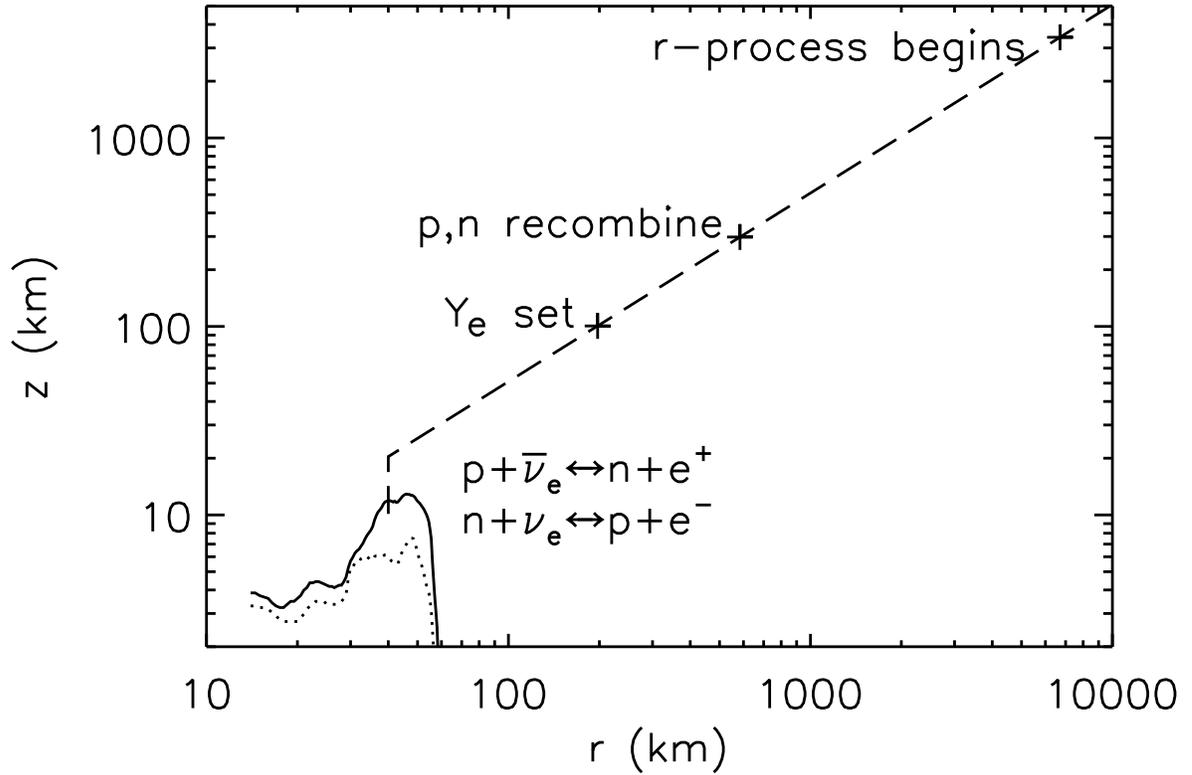}
\caption{Shows a schematic representation of the neutrino and antineutrino surfaces, for the
same vertical slice shown in Fig.\ \ref{fig:vertical} as well as an outflow trajectory, 
$\beta=0.8$, $s=20$.  Indicated on the figure are the relative positions where the
electron fraction is finished being set, where neutrons and protons recombine to begin to
produce nuclei, and where the final stage of nucleosynthesis, the $r$-process, takes place.
\label{fig:schematic}}
\end{figure}

Starting from this model of the disk, we proceed to calculate nucleosynthesis products in material
outflowing the disk.  We construct parameterized trajectories using the method outlined in
\citet{sur05}.  We take the outflow to be adiabatic with entropy per baryon $s$ in units of
Boltzmann's constant and velocity as a function of distance from the black hole $r$ to be
$v=v_{\infty}(1-r_{0}/r)^{\beta}$, where $r_{0}$ is the starting position on the disk, $v_{\infty}$ is
the final coasting velocity of $0.1c$, and $\beta$ determines the outflow acceleration.  Lower $\beta$
corresponds to a more rapidly accelerating outflow while higher $\beta$ is more slowly accelerating. 
This parameterization is qualitatively similar to numerical outflows models such as those in
\citet{bar08}.  Fig.\ \ref{fig:schematic} shows a sample trajectory for outflow from the neutrino
surface.  We expect the direction of the outflow to be vertical close to the disk and radially outward
far from it.  Therefore in constructing our trajectories we take the outflow to be vertical initially,
switching over to radial at a distance of twice the density scale height above the disk (though we
note the resulting nucleosynthesis is not particularly sensitive to the choice of turnover point).  As
can be seen from this figure, the electron fraction is set by the weak rates relatively close to the
surfaces, but nucleosynthesis takes place considerably further out.  The dominant effect of the weak
rates is from neutrino and antineutrino capture.  The neutrino fluxes necessary to calculate these
rates are determined by integrating over the contribution from the neutrinos emitted from all points
on the neutrino surfaces.  We calculate the element synthesis using a nuclear statistical equilibrium
code, a charged particle reaction code, and finally an $r$-process network code as described in
\citet{sur06}.  Given the asymmetry apparent in Figs.\ \ref{fig:nusph} and \ref{fig:vertical}, it is
clear that the trajectories from different starting points will produce different nucleosynthesis
results.  Therefore, for a given starting radius $r_0$, we run the complete calculation eight times,
for starting points at eight different equally spaced angles around the disk.  We then take the eight
different nucleosynthesis outcomes and average them.  In doing so we assume that material is not
ejected preferentially on one particular side of the disk.  When hydrodynamic calculations of winds
from these disks become available, this averaging will no longer be necessary.

\section{Results and Discussion}

\begin{figure}
\plotone{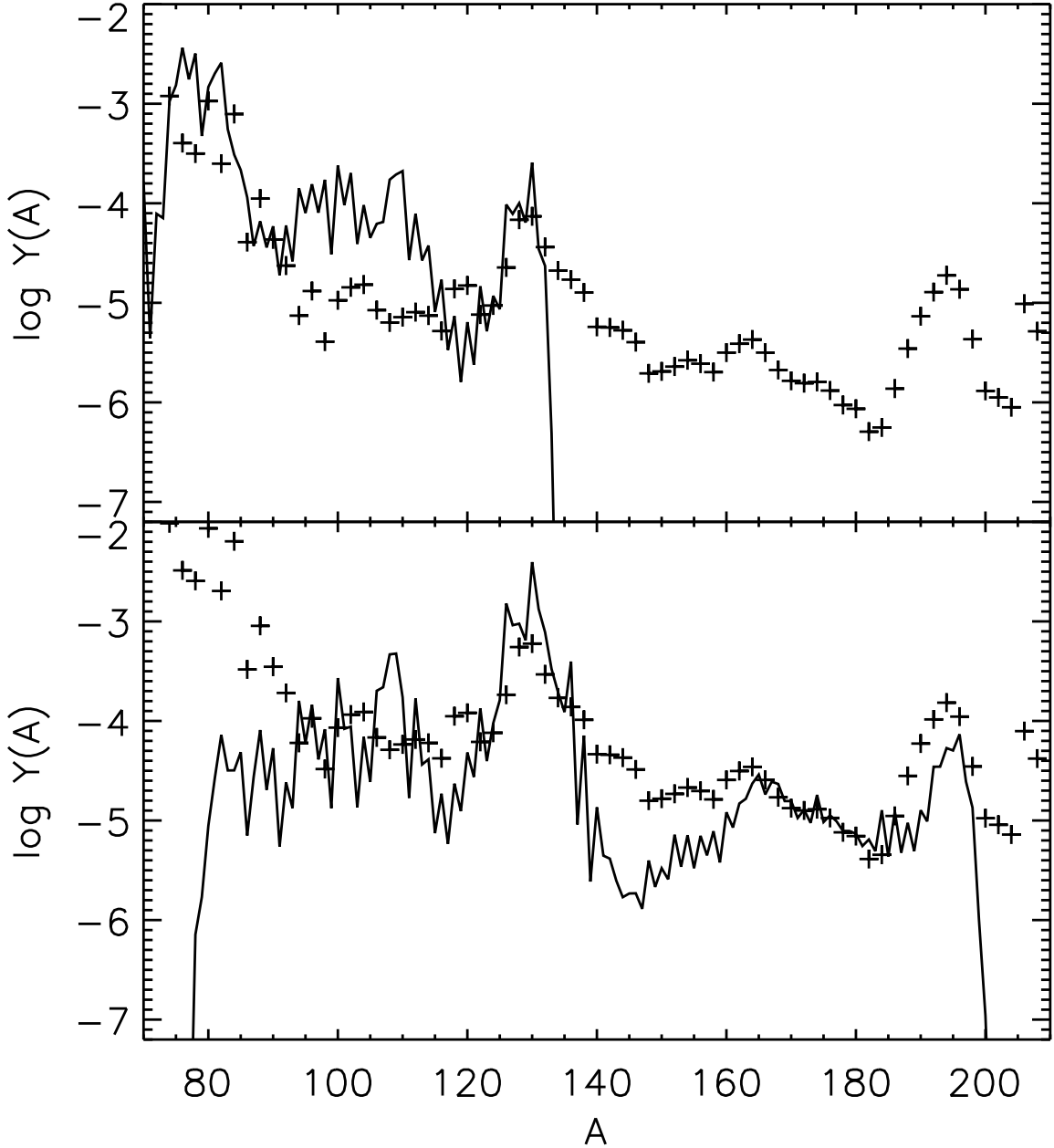}
\caption{Plots of the nucleosynthesis calculation for outflow from the disk.  The crosses
represent the scaled solar system $r$-process abundances.  The solid lines show the
calculation.  The top panel shows a weak $r$-process from conditions, $s=20$, $\beta=0.2$,
angle averaged over the disk. This result is typical of outflow conditions where $s < 30$
or $\beta < 0.8$.  The bottom panel shows a main $r$-process from the conditions, $s=40$,
$\beta = 0.8$,  again angle averaged around the disk.  This result is typical for conditions
in the range $s=30$ to $s=50$, $\beta \geq 0.8$.
\label{fig:rprocess}}
\end{figure}

We have undertaken a wide survey of the nucleosynthesis from a variety of trajectories with conditions
$0.2\leq \beta \leq 1.4$, $10 \leq s \leq 50$, and 20 km $\leq r_{0} \leq 80$ km. In choosing these
ranges we are guided by the results of \citet{pru04,met07,bar08} for neutrino- and pressure-driven
flows.  We have found our results can be divided into two categories: those that produce a weak
$r$-process and those that produce a main $r$-process.  Examples of each type are shown in Fig.\
\ref{fig:rprocess}.  While in this figure we have taken the average of the results for trajectories
with eight different starting points, we find $r$-process nuclei produced in all trajectories.

The $Y_e$ in the outflow is determined by the balance of the weak interactions.  In this case $\nu_e$ and
$\bar{\nu}_e$ capture eventually dominates over $e^+$ and $e^-$ capture because the former have higher
temperatures in the region above the torus.  The neutrino capture rates for the relevant energy scales are
approximately $\propto n_\nu T_\nu^2$, where $n_\nu$ is the neutrino number flux; for more detail see
\cite{mcl96}.  In the present merger scenario, not only is $T_{\bar{\nu}_e}$ substantially greater than
$T_{\nu_e}$, but $n_{\bar{\nu}_e} > n_{\nu_e}$ as well \citep{set06}.  Therefore in all cases $\bar{\nu}_e + p
\rightarrow n + e^+$ is faster than $\nu_e + n \rightarrow p + e^-$ above the disk and the outflow becomes
neutron rich \citep{mcl05}.  If these conditions were not met, the outflow could have become proton rich,
as can happen in other astrophysical environments \citep{fro06,bur06}.  The electron fractions for a
number of our trajectories at the point of nuclear recombination ($Y_{e,nr}$) and at the onset of the
$r$-process ($Y_{e,f}$) are given in Table \ref{tab:ye}.

Weak $r$-process nuclei are produced in low entropy or fast acceleration conditions, and a main
$r$-process is produced in between these two extremes.  In all cases we take the initial
electron fraction to be that given by the numerical model for the accretion torus.  However, the
weak interactions are sufficiently rapid that the electron fraction becomes reset in the
outflow.  When material first leaves the torus, the electron fraction spikes due to electron and
positron capture, then later as the neutrinos dominate, $Y_e$ falls.  Both outflow timescale
(i.e. $\beta$) and entropy have an impact on the electron fraction.  The primary effect of
outflow timescale is to influence the number of neutrino interactions that the nucleons can
have. In the case of fast acceleration, there is relatively little time for neutrino capture on
nucleons, so $Y_e$ does not fall as far, and only a weak $r$-process is produced.  For the
conditions considered here, the entropy has two effects: it determines the initial spike in the
electron fraction and it influences the point at which nuclei start to form.  In the case of low
entropy, nuclei form early and so the alpha effect, see e.g. \citet{mcl96}, which drives up the
electron fraction, is strong. This prevents the production of a full $r$-process.  We note that
while decreasing the timescale will weaken the alpha effect by allowing fewer neutrino captures
during the epoch of alpha particle formation, it also decreases the number of neutrino captures
before alpha particle formation.  This means that the electron fraction never becomes
sufficiently low to create a main $r$-process. Thus it is only for high entropy ($s\geq 30$) and
moderately accelerating trajectories ($\beta \geq 0.8$) that a main $r$-process occurs.

\section{Conclusions}

We find that for BH-NS mergers, almost all possible wind trajectories from the inner regions
of the disk result in at least a weak $r$-process.  This result is directly attributable to
the neutrino physics.  Should the connection between short duration GRBs and BH-NS mergers
become firmly established, our results imply that short duration GRBs produce $r$-process
elements.

In order for NS-NS mergers to be the sole contributor to the solar system weak $r$-process
abundances, then $10^{-1} \, {\rm M_\odot}$ to $1 \, {\rm M_\odot}$ must be ejected per
merger, given a rate of $10^{-4}$ to $10^{-5}$ per year in the Galaxy \citep{bel07}.  The
rate of BH-NS mergers is less well constrained.  However, given that current estimates of
mass ejected from compact object mergers tend to be not more than $10^{-2} \, {\rm M}_\odot$
\citep{ruf97,oec07}, unless the rate is higher there is additional source of weak $r$-process
material beyond BH-NS mergers. In any event the contribution of BH-NS mergers should be
included in the Galactic tally of $r$-process abundances.  Furthermore, if recent
predictions for bimodal production processes for compact object mergers are confirmed
\citep{bel06}, then there may be a contribution to the $r$-process early in the evolution of
the Galaxy.  This could influence the interpretation of the $r$-process pattern seen in halo
star data.

Given that an $r$-process occurs in the hot outflows over such a wide range of our
parameter study, a next step in determining the complete $r$-process abundance pattern
produced by BH-NS mergers would be to add the calculations presented here together with
estimates of the $r$-process from material ejected tidally during the merger.  This requires
future hydrodynamical simulations.

\acknowledgments

This work was partially supported by the Department of Energy under contracts
DE-FG05-05ER41398 (RS) and DE-FG02-02ER41216 (GCM). This work was partially supported by the
United States National Science Foundation under contract PHY-0244783 and AST-0653376 (WRH).
Oak Ridge National Laboratory (WRH) is managed by UT-Battelle, LLC, for the U.S. Department
of Energy under contract DE-AC05-000R22725.

\clearpage

\begin{deluxetable}{cccc} 
\tablewidth{0pt}
\tablecaption{Electron fractions for a sample of trajectories with $r_{0}=40$ km.}
\tablehead{\colhead{S} & \colhead{$\beta$} & \colhead{$Y_{e,nr}$} & \colhead{$Y_{e,f}$}}
\startdata
10 & 0.8 & 0.14 - 0.19 & 0.32 - 0.36 \\
10 & 1.4 & 0.14 - 0.19 & 0.39 - 0.42 \\
40 & 0.8 & 0.23 - 0.29 & 0.27 - 0.31 \\
40 & 1.4 & 0.17 - 0.19 & 0.23 - 0.26 \\
\enddata
\label{tab:ye}
\end{deluxetable}

\clearpage

\end{document}